\documentstyle[prd,aps,preprint,floats,epsf,rotate]{revtex}
\begin{document}
\draft
\title
{\bf \LARGE Self-dual Maxwell Chern-Simons Solitons In 1+1 Dimensions}
\narrowtext
\author{Prasanta K. Tripathy\thanks{e-mail: prasanta@iopb.stpbh.soft.net}}
\address{Institute of Physics, Bhubaneswar 751005, India}
\maketitle

\begin{abstract}
We study the domain wall soliton solutions in the relativistic self-dual
Maxwell Chern-Simons model in $1+1$ dimensions obtained by the dimensional 
reduction of the $2+1$ model. Both topological and nontopological self-dual 
solutions are found in this case. {\it A la} BPS dyons here the Bogomol'ny 
bound on the energy is expressed in terms of two conserved quantities.
We discuss the underlying supersymmetry. Nonrelativistic limit of this model 
is also considered and static, nonrelativistic self-dual soliton solutions 
are obtained.
\end{abstract}
\pacs{PACS numbers : 11.27.+d,11.10.Kk }

%\begin{multicols}{2}
\narrowtext

\newpage
\section{INTRODUCTION}
\hspace*{.8cm} The $2+1$ dimensional Maxwell Chern-Simons (MCS) system has 
already been studied, and existence of charged vortices of finite energy 
has been shown \cite{samir}. Self-dual topological and nontopological 
soliton solutions can be obtained in this case provided one also adds a
neutral scalar field to the theory\cite{clee90}. 
The nonrelativistic limit of this model was also considered and the self-dual
soliton solutions have been obtained\cite{dunne91}.\\
\hspace*{.8cm} 
Some time ago the $1+1$ dimensional nonlinear Sigma model \cite{abrahm} 
were obtained by dimensional reduction of certain $2+1$ dimensional nonlinear
Sigma models and soliton solutions in the $1+1$ dimensional case were shown
to be similar to the BPS dyons of $3+1$ dimensional Yang-Mills Higgs theory. 
This work was extended further by the inclusion of the Chern-Simons term 
\cite{pijush} and self-dual soliton solutions were again obtained. 
Recently the $1+1$ dimensional reduction of the abelian Higgs model with pure 
Chern-Simons term was considered and explicit topological and nontopological
domain wall solutions were obtained.\\
\hspace*{.8cm}
The purpose of this paper is to consider the dimensional reduction of the 
abelian Higgs model with Chern-Simons (as well as Maxwell) term 
(and a neutral scalar field).  
We show the existence of the self-dual topological as well as nontopological
domain wall solutions in this dimensionally reduced theory. The nonrelativistic
limit of this model is also considered and soliton solutions are again obtained.
 In section II we obtain the model by dimensional reduction of the $2+1$ 
MCS model. Here we study the invariance of the Lagrangian and the 
corresponding two conserved charges. In section III we obtain the BPS-type 
bound \cite{bps} 
on energy and show that the bound is saturated when the self-dual equations are
satisfied by the fields. In section IV we study the self-dual equations and 
obtain both the topological and the nontopological domain wall solutions for
the system. In section V we consider the underlying $N=2$ supersymmetry.
In section VI we consider the nonrelativistic limit of the model 
and again obtain the self-dual soliton solutions and study their properties.
Finally in section VII we conclude the results.

\section{THE MODEL}
The Lagrangian for the Maxwell-Chern Simons system is given by \cite {clee90,dunne91}
\begin{eqnarray}
{\cal L}_{2+1}^{MCS} &=&-\frac{1}{4e^{2}}F_{\rho\nu}F^{\rho\nu}
+\frac{\mu}{2e^{2}}\epsilon^{\eta\nu\rho}A_{\eta}\partial_{\nu}A_{\rho}
+(D_{\rho}\phi)^{\ast}(D^{\rho}\phi)\nonumber\\
&+&\frac{1}{2e^{2}}\partial_{\rho}N
\partial^{\rho}N 
-\frac{1}{c^{2}}|\phi|^{2}\left(N-\frac{e^{2}v^{2}}{{\mu}c}\right)^{2}
-\frac{e^{2}}{2c^{2}}\left(|\phi|^{2}-\frac{{\mu}c}{e^2}N\right)^{2}
\end{eqnarray}
Here $N$ is a real scalar field, $\phi$ is complex scalar field,
$c$ is the velocity of light and $A_{\mu}$s are gauge fields. 
This model has two degenerate vacua. The symmetric phase, 
having vacuum expection value $<\phi>=0,<N>=0$ and the asymmetric phase having 
$<\phi>=v,<N>=\frac{e^{2}v^{2}}{{\mu}c}$. In the symmetric phase the complex
scalar field $\phi$ has mass $\frac{e^{2}v^{2}}{{\mu}c^{2}}$ . The neutral 
scalar field $N$ and the gauge fields have the mass $\mu $. In asymmetric phase
there are two massive gauge degrees of freedom with masses given by\cite{khare}
\begin{eqnarray}
m_{\pm}^{2}=\frac{2e^{2}v^{2}}{c^{2}}+\frac{{\mu}^{2}}{2}\pm \frac{\mu}{2}
\sqrt{{\mu}^{2}+\frac{8e^{2}v^{2}}{c^{2}}}
\end{eqnarray}
The scalar fields also combine into two massive modes with masses $m_{\pm}$.
This model (in $2+1$ dimensions) possesses a Bogomol'ny-type 
bound \cite{bogom} and has static self-dual soliton solutions.\\
\hspace*{.8cm}
After compactification of the $y$ direction we get the following Lagrangian in 
$1+1$ dimensions,
\begin{eqnarray} 
{\cal L}_{1+1}^{MCS}&=&\frac{1}{2e^{2}}F_{01}^{2}
+\frac{{\mu}}{e^{2}}R(x)F_{01}(x)+
(D_{\rho}\phi)^{\ast}(D^{\rho}\phi)+\frac{1}{2e^{2}}\partial_{\rho}N
\partial^{\rho}N \nonumber \\
&&+\frac{1}{2e^{2}}\partial_{\rho}R\partial^{\rho}R-U(R,\phi,N) 
\end{eqnarray}
where
\begin{eqnarray}
&&D_{\rho}=\partial_{\rho}-iA_{\rho}\nonumber\\
&&U(R,\phi,N)=R^{2}(x)|\phi|^{2}+|\phi|^{2}\left(N
-\frac{e^{2}}{{\mu}}v^{2}\right)^{2}
+\frac{e^{2}}{2}\left(|\phi|^{2}-\frac{{\mu}}{e^{2}}N\right)^{2}.
\end{eqnarray}
Here for simplicity we put $c=1$.
We identify the $y$ independent component of $A_{y}(t,x,y)$ as $R(x)$.  
The symmetric phase is again given by $<\phi>=0 , <N>=0$, but now $<R>=R_{0}$ 
where $R_{0}$ is arbitrary. 
In this case the gauge fields as well as the $R$ field are massless  
while the $N$ field has mass ${\mu}$ and the Higgs field has mass 
$\sqrt{R_{0}^{2}+\frac{e^{4}v^{4}}{{\mu}^{2}}}$. 
In the broken phase $<N>=\frac{e^{2}v^{2}}{{\mu}},<\phi>=v,<R>=0$.
In this case the gauge fields and the scalar field R become massive
having masses equal to $\sqrt{2e^{2}v^{2}}$, while the scalar field $N$ 
and Higgs field combine to give two massive modes with masses $m_{\pm}$ 
given by eq.(2) with $c=1$.\\
\hspace*{.8cm} We express the Lagrangian in terms of dimensionless fields 
$n(\tilde x), r(\tilde x), f_{01}(\tilde x)$ and $\varphi(\tilde x)$ where 
\begin{eqnarray} 
N(x)&=&\frac{e^{2}v^{2}}{\mu}n(\tilde x),\hspace{.5cm} 
R(x)=\frac{e^{2}v^{2}}{\mu}r(\tilde x), \nonumber \\
\phi(x)&=&v\varphi(\tilde x), \hspace{.85cm}
A_{\rho}(x)=\frac{e^{2}v^{2}}{\mu}a_{\rho}(\tilde x),\\
x^{\mu}&=&\frac{\mu}{e^{2}v^{2}}\tilde x^{\mu} 
\end{eqnarray}
From now onwards we do the calculations with the new field variables which are
functions of $\tilde x^{\mu}$. We shall omit tilde from $x$ and now we will
take $x$ for $\tilde x$ unless otherwise specified.  
Then the dimensionless Lagrangian becomes,
\begin{eqnarray}
\frac{{\mu}^{2}}{e^{4}v^{6}}{\cal L}_{1+1}^{MCS} = L&=&\frac{1}{2k}f_{01}^{2}(x)
+r(x)f_{01}(x)+(d_{\rho}\varphi)^{\ast}(d^{\rho}\varphi)\nonumber \\
&&+\frac{1}{2k}\partial_{\rho}n\partial^{\rho}n+ 
\frac{1}{2k}\partial_{\rho}r\partial^{\rho}r-V(f,n,r)
\end{eqnarray} 
where,
\begin{eqnarray} 
&&V(f,n,r)=f(x)r^{2}(x)+f(x)\left(n(x)-1\right)^{2}+
\frac{k}{2}\left(f(x)-n(x)\right)^{2}\nonumber\\
&&f(x)=|\varphi|^{2},\hspace{.5cm}k=
\frac{{\mu}^{2}}{e^{2}v^{2}},\hspace{.5cm}
d_{\rho}=\partial_{\rho}-ia_{\rho}(x),
\end{eqnarray}
In the limit $k\longrightarrow\infty $ the Lagrangian reduces to\cite{hkao}
\begin{equation}
L=r(x)f_{01}(x)+(d_{{\rho}}\varphi)^{\ast}(d^{\rho}\varphi)-
f(x)r^{2}(x)
-f(x)(f(x)-1)^{2}
\end{equation}
which is the pure Chern-Simons Higgs Lagrangian. This system admits known 
topological and nontopological soliton solutions.
The theory is invariant under $U(1)$ gauge transformation and the 
corresponding current is, 
\begin{eqnarray}
j^{\rho}=i\{(d^{\rho}\varphi)^{\ast}\varphi-
\varphi^{\ast}d^{\rho}\varphi\}
\end{eqnarray}
\vspace{.2in}
The theory has another invariance. In particular there exists another set of transformations of fields 
\begin{eqnarray}
&&\delta\varphi(x)=i\varphi\left(r(x)+\frac{1}{k}f_{01}(x)\right),\hspace{.5cm} 
\delta a_{1}(x)=j^{0}+\frac{(n^{2}+r^{2})'}{2\left(r(x)
+\frac{1}{k}f_{01}(x)\right)}, \nonumber\\
&&\delta a_{0}(x)=j^{1}+\frac{C_{0}}{-r(x)+\frac{f_{01}(x)}{k}},\hspace{.5cm}
\delta n(x)=0,\hspace{.5cm}\delta r(x)=0.
\end{eqnarray}
where prime denotes derivative with respect to space coordinate.
Here $C_{0}$ is some arbitrary constant.
The above transformations leave the theory invariant only after using the
equations of motion.The corresponding charge is :
\begin{eqnarray}
\left(\frac{\mu}{e^{2}v^{4}}\right)Y
=\int dx \left\{(d^{0}\varphi)^{\ast}\delta\varphi+\delta\varphi^{\ast}
d^{0}\varphi+(r(x)+\frac{f_{01}(x)}{k})\delta a_{1}(x)\right\}
\end{eqnarray}

\section{Equation Of Motion And Self duality}
Now varying the Lagrangian(7) with respect to the fields we get 
the equations of motion:
\begin{eqnarray}
&&\frac{1}{k}f^{\prime}_{01}(x)+r^{\prime}(x)
-i\left((d_{0}\varphi)^{\ast}\varphi
-\varphi^{\ast}d^{0}\varphi\right) = 0,\\
&&\frac{1}{k}\partial_{\rho}\partial^{\rho}r+2r(x)f(x)-f_{01}(x)=0,\\
&&-\frac{1}{k^2}\partial_{\rho}\partial^{\rho}n+\left(f(x)-n(x)\right)
-\frac{2}{k}f(x)\left(n(x)-1\right)=0,\\
&&d_{\rho}d^{\rho}\varphi+\{r^{2}(x)+\left(n(x)-1\right)^{2}
+k\left(f(x)-n(x)\right)\}\varphi=0.
\end{eqnarray}
where eq.(13) is the Gauss law equation. Using the Gauss law equation,
(and putting appropriate dimensions )
Q and Y as given by eqs.(10) and (12) take the form,
\begin{eqnarray}
&&Q=ev^{2}\int_{-\infty}^{+\infty}j_{0}(x)dx,\\
&&Y=\left(\frac{e^{2}v^{4}}{\mu}\right)\int_{-\infty}^{+\infty}dx\frac{1}{2}
\left\{n^{2}(x)+r^{2}(x)\right\}'
\end{eqnarray}

The dimensionless energy density ${\cal E}(x)$ is given by
\begin{eqnarray}
\frac{{\mu}^{2}}{v^{6}e^{4}}\tilde{\cal E}(x)={\cal E}(x)
&=&\frac{1}{2k}\left(f_{01}^{2}(x)
+r^{\prime2}(x)+n^{\prime2}(x) 
+\dot{r}^{2}(x) +\dot{n}^{2}(x)\right) \nonumber \\
&&+|d_{0}\varphi|^{2}+|d_{1}\varphi|^{2}+V(f,n,r)
\end{eqnarray}
Here overdot denotes derivative with respect to time. We study the system  
in the static ansatz where we have $\dot{n}(x)=0=\dot{r}(x)$.  
Integrating by parts and using the Gauss law equation the energy density 
can be expressed as 
\begin{eqnarray}
{\cal E}(x)&=&|d_{0}\varphi+i(n(x)-1)\varphi(x)\cos\alpha
-ir(x)\varphi(x) \sin\alpha|^{2} \nonumber\\
&&+|d_{1}\varphi+(n(x)-1)\varphi(x)\sin\alpha
+r(x)\varphi(x)\cos\alpha|^{2} \nonumber \\
&&+\frac{1}{2k}\left(r^{\prime}(x)+k\left(f(x)-n(x)\right)\cos \alpha
-f_{01}(x)\sin\alpha\right)^{2}\nonumber\\
&&+\frac{1}{2k}\left(n^{\prime}(x)+k\left(f(x)-n(x)\right)\sin\alpha
+f_{01}\cos\alpha\right)^{2} \nonumber \\
&&+\frac{1}{2}\left(n^{2}(x)+r^{2}(x)\right)^{\prime}\sin\alpha
+\left(r(x)\left(1-f(x)\right)\right)^{\prime}\cos\alpha \nonumber \\
&&+\left(\frac{1}{k}f_{01}(x)\left(r(x)\sin\alpha +\left(1-n(x)\right) 
\cos\alpha\right)+f(x)\left(1-n(x)\right)\sin\alpha\right)'\nonumber \\
\end{eqnarray}
Here $\alpha$ is an arbitrary angle variable.
The last term in the above expression vanishes after integration as
$f_{01}(x) \longrightarrow0$ asymptotically.
Then we can write the following inequality:
\begin{equation}
{\cal E}(x)\geq\frac{1}{2}\left(n^{2}(x)+ r^{2}(x)\right)^{\prime}\sin\alpha
+\left(r(x)(1-f(x))\right)^{\prime}\cos\alpha
\end{equation}
This is a Bogomol'ny type bound.
The lower bound on energy is saturated when the 
following self-dual equations hold
\begin{eqnarray}
&&d_{0}\varphi+i(n(x)-1)\varphi(x)\cos\alpha-ir(x)\varphi(x)\sin\alpha=0,
\nonumber\\
&&d_{1}\varphi+(n(x)-1)\varphi(x)\sin\alpha+r(x)\varphi(x)\cos\alpha=0,
\nonumber\\
&&\frac{1}{k}(r^{\prime}(x)-f_{01}(x)\sin\alpha)+(f(x)-n(x))\cos\alpha=0,
\nonumber\\
&&\frac{1}{k}(n^{\prime}(x)+f_{01}(x)\cos{\alpha})+(f(x)-n(x))\sin\alpha=0.
\end{eqnarray}
These along with the Gauss law eq.(13)
are consistent with the second order static equations of motion as given by
eqs.(14) to (16).
We can rewrite the above self-dual equations as,
\begin{eqnarray}
&&\frac{1}{k}f^{\prime}_{01}(x)+r^{\prime}(x)
+2f(x)\{(n(x)-1)\cos\alpha-r(x)\sin\alpha\}=0,\nonumber\\
&&f^{\prime}(x)+2f(x)\{(n(x)-1)\sin\alpha+r(x)\cos\alpha\}=0,\nonumber\\
&&\frac{1}{k}(r^{\prime}(x)-f_{01}(x)\sin\alpha)+(f(x)-n(x))\cos\alpha=0,
\nonumber\\
&&\frac{1}{k}(n^{\prime}(x)+f_{01}(x)\cos{\alpha})+(f(x)-n(x))\sin\alpha=0.
\end{eqnarray}
Eliminating the fields $r, f_{01} $and $n$ from the above equations we find
the fourth order equation:
\begin{eqnarray}
-\frac{1}{k^{2}}(\ln f)''''+\left(1+\frac{2f(x)}{k}\right)(\ln f)''
+\frac{2}{k}f''(x)-4f(x)(f(x)-1)=0.
\end{eqnarray}
\hspace*{.8cm}
In the self-dual limit the energy
\begin{equation}
E=\frac{e^{2}v^{4}}{\mu}\int_{-\infty}^{+\infty}dx{\cal E}(x)
\end{equation}
can be writen as
\begin{equation}
E=\left\{Y\sin\alpha+\left(\frac{ev^{2}}{\mu}\right)Q\cos\alpha\right\}
\end{equation}
where we have used the fact that $(rf)$ vanishes at both $\pm\infty$.

\section{Asymptotic Properties}
The finiteness of energy requires that the fields have to satisfy some 
boundary conditions. From the expression of energy density we find that the
fields have to take one of the following two set of values as 
$x\rightarrow \pm\infty$. They are:
\begin{eqnarray}
&& f(x)=1, \hspace{.5cm} n(x)=1, \hspace{.5cm}
f_{01}(x)=0, \hspace{.5cm} r(x)=0  .
\end{eqnarray}
or
\begin{eqnarray}
&& f(x)=0, \hspace{.5cm} n(x)=0, \hspace{.5cm}
f_{01}(x)=0, \hspace{.5cm} r(x)=r_{0} .
\end{eqnarray}
Where $r_{0}$ is an arbitrary real constant.
For the asymmetric solutions the fields at $+\infty$ take one of the above
sets of values and at $-\infty$ the other. But for the symmetric solutions
the fields takes the second set of values both at $+\infty$ and at $-\infty$.\\
\hspace*{.6cm}
\subsection{Asymmetric Solution}
It is difficult to solve the self-dual equations analytically.
However we can obtain the asymptotic form of the various fields. 
Consider first the asymmetric solution for which let the fields take the values
(27) at $-\infty$ and (28) at $\infty$. Then 
from the fourth order equation we find that for $x\longrightarrow -\infty$,
the field $f$ behaves as
\begin{eqnarray}
f(x)\longrightarrow(1-qe^{lx})+....,
\end{eqnarray}
where 
\begin{eqnarray}
l=-\frac{k}{2}+\frac{1}{2}\sqrt{k^{2}+8k}.
\end{eqnarray}
Here $q$ is some arbitrary constant. \\
Using the above expression for $f$ in the self dual equations we find that 
the behaviour of the fields $n, r$ and $f_{01}$ as 
$x\longrightarrow -\infty$ are, 
\begin{eqnarray}
&&n(x)\longrightarrow(1-q \frac{l}{2}e^{lx})+....  ,\nonumber\\
&&r(x)\longrightarrow q\frac{l\cos\alpha}{2(1-\sin\alpha)}e^{lx}+....  ,\nonumber\\
&&f_{01}(x)\longrightarrow q\frac{l^{2}\cos\alpha}{2(1-\sin\alpha)}e^{lx}+.... .
\end{eqnarray}
Similarly for $x\longrightarrow \infty$ we find, from the fourth order equation 
\begin{eqnarray}
f(x)\longrightarrow a_{0}e^{-mx}\left(1+\tilde{a}_{0}e^{-mx}\right)+..... ,
\end{eqnarray}
where
\begin{eqnarray}
-\frac{m^{4}}{k^{2}}\frac{\tilde{a}_{0}}{a_{0}}
+m^{2}\left(\frac{\tilde{a}_{0}}{a_{0}}+\frac{2}{k}\right)+4=0
\end{eqnarray}
Then from the self-dual equations we find,
\begin{eqnarray}
&&n(x)\longrightarrow b_{0}e^{-mx}+..... ,\nonumber\\
&&r(x)\longrightarrow \tilde d_{0}+d_{0}e^{-mx}+..... ,\nonumber\\
&&f_{01}(x)\longrightarrow c_{0}e^{-mx}+..... .
\end{eqnarray}
where the coefficients $a_{0}, b_{0}, c_{0}, d_{0} $ and $\tilde d_{0}$  
satisfy the following relations,
\begin{eqnarray}
&&\frac{b_{0}}{a_{0}}=\frac{k+2}{k^{2}-m^{2}},\nonumber\\
&&\frac{c_{0}}{a_{0}}=\frac{m(2+m\sin\alpha)
+k(m+2\sin\alpha)}{(k^{2}-m^{2})\cos\alpha},\nonumber\\
&&\frac{d_{0}}{a_{0}}=\frac{k(2+m\sin\alpha)
+m(m+2\sin\alpha)}{m(m^{2}-k^{2})\cos\alpha},\nonumber\\
&&\tilde d_{0}=\frac{m+2\sin\alpha}{2\cos\alpha}.
\end{eqnarray}
The charges $Q$ and $Y$ are given in this case by 
(putting appropriate dimensions)
\begin{eqnarray}
Y&=&\left(\frac{e^2v^4}{\mu}\right)\left(\frac{\tilde{d}_{0}^{2}-1}{2}\right)\\
Q&=&ev^2\tilde{d}_{0}  .
\end{eqnarray}
while $E$ is given by eq.(26). In case one assumes the same exponent 
at $\pm \infty$ i.e. $m=l$ then $Y$, $Q$ and $E$ are solely determined 
by $k$ and the angle $\alpha$. Further as $k\longrightarrow\infty$, $Y$, $Q$
and $E$ reduces to their expressions in the pure Chern-Simons case\cite{hkao}.

We have also found numerical solution to the fourth order eq.(24) in the
asymmetric case with the boundary values as given by eqs.(29) and (32).
The profile of the Higgs field in case $k=0.01$ is given in fig.$1$. 
%%%%%%%%%%%%%%%%%%%%%%%%%%%%%%%%%%%%%%%%%%%%%%%%%%%%%%%%%%%%%%%%
\begin{figure}[ht]
\begin{center}
\vskip -1 in
\leavevmode
\epsfysize=12truecm \vbox{\epsfbox{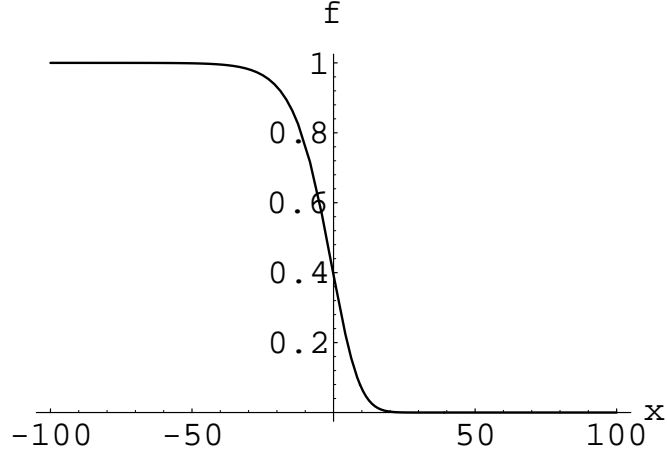}}
\vskip -1in
\end{center}
\caption{ Asymmetric Solution for k=0.01}
\end{figure}
%%%%%%%%%%%%%%%%%%%%%%%%%%%%%%%%%%%%%%%%%%%%%%%%%%%%%%%%%%%%%%%%%%
\newpage

\subsection{Symmetric Solution}
For the symmetric case, we assume that the profile of the field $f$ is 
symmetric around $x=0$. To know the behaviour of $f$ around $x=0$, we
expand the field $f$ as,
\begin{eqnarray}
f(x)= a+bx^{2}+cx^{4}+.....
\end{eqnarray}
On substituting the above expression in eq.(24) we get
\begin{eqnarray}
-6c+\left(\frac{k}{2}+2a\right)kb-a^{2}k^{2}(a-1)=0
\end{eqnarray}
For $x\longrightarrow \infty$ we find from the fourth order equation,  
that $f(x)$ is similar to that given in eq.(32), and hence the fields
$n, r$ and $f_{01}$ are similar to those given in eq.(34).\\
\hspace*{.8cm}
The expressions for $Y, Z$ and $E$ for the symmetric domain wall soliton 
are found to be (putting appropriate dimensions)
\begin{eqnarray}
&&Y=\frac{1}{2}\left(\tilde d_{2}^{2}-\tilde d_{1}^{2} \right)
=\frac{e^2v^4}{\mu}\left(\frac{m\sin\alpha}{\cos^{2}\alpha}\right),\\
&&Q=ev^2\left(\tilde d_{2}-\tilde d_{1}\right)  
=\frac{mev^2}{\cos\alpha},\\
&&E= \frac{e^2v^4}{\mu}\left(\frac{m}{\cos^{2}\alpha}\right).
\end{eqnarray}
where $\tilde{d}_{1}$ and $\tilde{d}_{2}$ are the value of $r(x)$ 
at $-\infty$ and at $+\infty$ respectively.  
From the above expressions it can be seen that 
\begin{eqnarray}
E=\sqrt{Y^{2}+\left(\frac{ev^2}{\mu}Q\right)^{2}}
\end{eqnarray}
and $Y=E\sin\alpha , \frac{ev^2}{\mu}Q=E\cos\alpha$ .
Then the ratio of energy to charge is given by
\begin{eqnarray}
\frac{E}{Q}&=&\frac{ev^{2}}{{\mu}\cos\alpha}\\
&=&\frac{1}{e}\sqrt{\overline{R}^{2}+\frac{e^{4}v^{4}}{{\mu}^{2}}}
\end{eqnarray}
where $\overline R=\tan\alpha$ is the average value of $R$.
Mass of the elementary excitation in the unbroken phase is 
$m_{e}=\sqrt{{R}_{0}^{2}+\frac{e^{4}v^{4}}{{\mu}^{2}}}$. 
Taking $R_{0}=\overline R$ we find 
\begin{eqnarray}
\frac{E}{Q}=\frac{m_{e}}{e}
\end{eqnarray}
which means that the symmetric solution is at the threashold of stability.\\
\hspace*{.8cm}
We have also found numerical solutions to the fourth order eq.(24) in the 
symmetric case with the boundary value as given by eqs.(40) and (32). 
The profile of the Higgs field $f(x)$ is given in case $k=.01$ (we have chosen
$a=0.8, b=0.08,$ where $a, b$ are given by eq.(39)).

%%%%%%%%%%%%%%%%%%%%%%%%%%%%%%%%%%%%%%%%%%%%%%%%%%%%%%%%%%%%%%%%
\begin{figure}[ht]
\begin{center}
\vskip -1 in
\leavevmode
\epsfysize=12truecm \vbox{\epsfbox{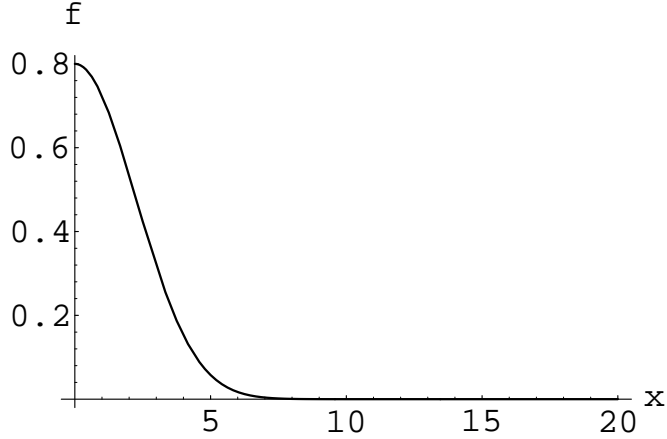}}
\vskip -1in
\end{center}
\caption{ Symmetric Solution for k=0.01 }
\end{figure}
%%%%%%%%%%%%%%%%%%%%%%%%%%%%%%%%%%%%%%%%%%%%%%%%%%%%%%%%%%%%%%%%%%
\newpage
\section{Supersymmetry}
We have obtained the supersymmetric model by dimensional reduction of the 
three dimensional supersymmetric Maxwell Chern-Simons model
\cite{hmin,bhlee}. The Lagrangian 
in two dimensions is found to be
\begin{eqnarray}
L=&& \frac{1}{2e^2}F_{01}^{2}
     +\frac{\mu}{e^2}R F_{01}
     +\frac{1}{2e^2}\partial_{\rho}R\partial^{\rho}R\nonumber\\
  && +\frac{1}{2e^2}\partial_{\rho}N\partial^{\rho}N
     +(D_{\rho}\phi)^{\ast}(D^{\rho}\phi)
     -R^{2}|\phi |^{2}\nonumber\\
  && -\frac{e^2}{2}\left(|\phi |^{2}+\frac{\mu}{e^{2}}N-v^{2}\right)^{2}
     -N^{2}|\phi |^{2}\nonumber\\
  && +i\overline{\psi} \gamma^{\rho} D_{\rho}\psi
     -i\overline{\psi} \gamma^{5} R \psi
     +\frac{i}{e^2} \overline{\chi} \gamma^{\rho} \partial_{\rho}\chi\nonumber\\
  && -\frac{\mu}{e^2} \overline{\chi}\chi
   +i\sqrt{2}\left(\overline{\psi}\chi\phi-\overline{\chi}\psi\phi^{\ast}\right)
  -N\overline{\psi}\psi
\end{eqnarray}
The bosonic part of this Lagrangian is equal to eq.(3) after a redefinition
of the field $N$ by $\left(-N+\frac{e^{2}v^{2}}{\mu}\right)$.
The Lagrangian has the following $N=2$ supersymmetry.
\begin{eqnarray}
&&\delta_{\eta}A_{\rho}=
i\left(\overline{\eta}\gamma_{\rho}\chi-\overline{\chi}\gamma_{\rho}\eta\right)
\nonumber\\
&&\delta_{\eta}R=
\left(\overline{\chi}\gamma^{5}\eta-\overline{\eta}\gamma^{5}\chi\right)
\nonumber\\
&&\delta_{\eta}\phi=\sqrt{2}\overline{\eta}\chi\nonumber\\
&&\delta_{\eta}N=i\left(\overline{\chi}\eta-\overline{\eta}\chi\right)
\nonumber\\
&&\delta_{\eta}\psi=\sqrt{2}\left(-i\gamma^{\rho}\eta D_{\rho}\phi
+i\gamma^{5}\eta R\phi-\eta N\phi\right)\nonumber\\
&&\delta_{\eta}\chi=-i\left(\gamma^{0}\partial_{1}R+
\gamma^{1}\partial_{0}R\right)\eta+\gamma{5}F_{01}\eta
+\gamma^{\rho}\partial_{\rho}N\eta\nonumber\\
&&  -i\eta e^{2}\left(|\phi |^{2}+\frac{\mu}{e^2}N-v^{2}\right)
\end{eqnarray}
Here we have choosen $\gamma^{0}=\sigma_{2},\gamma^{1}=i\sigma_{1},
\gamma^{5}=\gamma^{0}\gamma^{1}=\sigma_{3}, \eta_{\rho\sigma}=diag(1, -1)$.
The supercharge which generates this transformation is given by
\begin{eqnarray}
Q=&&\int dx (\sqrt{2}(D_{\rho}\phi)^{\ast}\gamma^{\rho}\gamma^{0}\psi
    -\sqrt{2}R\phi^{\ast}\gamma^{1}\psi+\sqrt{2}iN\phi^{\ast}\gamma^{0}\psi
\nonumber\\
  &&-\frac{i}{e^2}\partial_{\rho}N\gamma^{\rho}\gamma^{0}\chi
    -\frac{i}{e^2}F_{01}\gamma^{1}\chi-\frac{1}{e^2}\gamma^{5}\partial_{0}R\chi
\nonumber\\
  &&+\frac{1}{e^2}\partial_{1}R\chi+\left(|\phi |^{2}+\frac{\mu}{e^2}N
    -v^{2}\right)\gamma^{0}\chi)
\end{eqnarray}
The superalgebra is found to be
\begin{eqnarray}
\frac{1}{2}\left\{Q_{\alpha},\overline{Q}_{\beta}\right\}
=\left(\gamma^{\rho}\right)P_{\rho}+\delta_{\alpha\beta}Z
 +i\left(\gamma^{5}\right)_{\alpha\beta}Y
\end{eqnarray}
Where the central charges $Y$, and $Z$ are given by,
\begin{eqnarray}
&&Y=\int dx \frac{\mu}{2e^2}\left(R^{2}
+\left(N-\frac{e^{2}v^{2}}{\mu}\right)^{2}\right)'\nonumber\\
&&Z=\int dx \left(R\left(|\phi |^{2}-v^{2}\right)\right)'
\end{eqnarray}
It can be easily varified that the central charges $Y$ and $Z$ are 
equal to the Noether and topological charges after a field redefinition.
\section{NONRELATIVISTIC MODEL}

To get the nonrelativistic Lagrangian we substitute
\begin{equation}
\phi=\sqrt{\frac{{\mu}c^{3}}{2e^{2}v^{2}}}
\left(e^{-imc^{2}t}\psi+e^{imc^{2}t}\tilde\psi\right)
\end{equation}
in Lagrangian (1).
Neglecting higher order terms of $\frac{1}{c}$ we have the nonrelativistic 
Lagrangian \cite{dunne91}
\begin{eqnarray}
{\cal L}_{2+1}^{NR}=&&-\frac{1}{4e^{2}}F_{{\rho}{\nu}}F^{{\rho}{\nu}}+
+\frac{\mu}{2e^{2}}\epsilon^{{\eta}{\nu}{\rho}}A_{\eta}\partial_{\nu}A_{\rho}
+\frac{1}{2e^{2}}\partial_{\rho}N\partial^{\rho}N\nonumber\\ 
&&-\frac{{\mu}^{2}}{2e^{2}}N^{2}+ic\psi^{\ast}D_{0}\psi
+ic\tilde\psi^{\ast}(\partial_{0}+iA_{0})\tilde\psi
-\frac{{\mu}c^{3}}{2e^{2}v^{2}}(D_{i}\psi)^{\ast}(D_{i}\psi)\nonumber\\
&&-\frac{{\mu}c^{3}}{2e^{2}v^{2}}(D_{i}\tilde\psi)^{\ast}(D_{i}\tilde\psi)
-\frac{{\mu}^{2}c^{4}}{8e^{2}v^{4}}\left(|\psi|^{2}+|\tilde\psi|^{2}\right)^{2}
+\left(1+\frac{{\mu}^{2}c^{2}}{2e^{2}v^{2}}\right)\left(|\psi|^{2}
+|\tilde\psi|^{2}\right)N
\end{eqnarray}
(Here all the fields are functions of $x^{\mu}$ and not of $\tilde x^{\mu}$). 
After dimensional reduction and restricting only to the zero anti-particle 
sector we get the following nonrelativistic Lagrangian in $1+1$ dimensions,
\begin{eqnarray}
{\cal L}_{1+1}^{NR}=&&\frac{1}{2e^{2}}F_{01}^{2}+\frac{{\mu}}{e^{2}}RF_{01}
+\frac{1}{2e^{2}}\partial_{\rho}R\partial^{\rho}R 
+\frac{1}{2e^{2}}\partial_{\rho}N\partial^{\rho}N
-\frac{{\mu}^{2}}{2e^{2}}N^{2}(x)+i\psi^{\ast}(x)D_{0}\psi\nonumber\\
&&-\frac{\mu}{2e^{2}v^{2}}(D_{x}\psi)^{\ast}(D_{x}\psi)
+\left(1+\frac{{\mu}^{2}}{2e^{2}v^{2}}\right)|\psi|^{2}N(x) 
-\frac{{\mu}^{2}}{8e^{2}v^{4}}|\psi|^{4}
-\frac{\mu}{2e^{2}v^{2}}R^{2}(x)|\psi|^{2}
\end{eqnarray}
This is same as the nonrelativistic limit of the dimensionally reduced 
relativistic $1+1$ dimensional Lagrangian. 
As before, let us express the Lagrangian in terms of the dimensionless fields 
which are functions of $\tilde x $, and omit tilde

\begin{eqnarray}
\frac{{\mu}^{2}}{e^{4}v^{6}}{\cal L}_{1+1}^{NR}
&=&L^{NR}=\frac{1}{2k}f_{01}^{2}+r(x)f_{01}(x)
+\frac{1}{2k}\partial_{\rho}r\partial^{\rho}r
+\frac{1}{2k}\partial_{\rho}n\partial^{\rho}n+2i\chi^{\ast}d_{0}\chi\nonumber\\
&&-(d_{x}\chi)^{\ast}(d_{x}\chi)
+2n(x)f(x)-\frac{k}{2}(f(x)-n(x))^{2}-r^{2}(x)f(x),
\end{eqnarray}
where $\chi = \frac{1}{ev}\sqrt{\frac{\mu}{2}} \psi $ is dimensionless.
The dimensionless energy density ${\cal E}(x)$ is 
\begin{eqnarray}
\frac{{\mu}^{2}}{e^{6}v^{4}}\tilde{\cal E}(x)
={\cal E}(x)&=&\frac{1}{2k}(f_{01}^{2}(x)+(r'(x))^{2}
+(n'(x))^{2}+\dot{r}^{2}(x)+\dot{n}^{2}(x))\nonumber\\
&&+\frac{k}{2}(n(x)-f(x))^{2}+(d_{x}\chi)^{\ast}(d_{x}\chi)
+r^{2}(x)f(x)-2n(x)f(x)
\end{eqnarray}
The equations of motion are
\begin{eqnarray}
&&2id_{0}\chi+d_{x}^{2}\chi+(2+k)n(x)\chi(x)-(kf(x)+r^{2}(x))\chi(x)=0,
\nonumber\\
&&\frac{1}{k}\partial_{\rho}\partial^{\rho}n+k(n(x)-f(x))-2f(x)=0,\nonumber\\
&&\frac{1}{k}\partial_{\rho}\partial^{\rho}r-f_{01}(x)+2r(x)f(x)=0,\\
&&\frac{1}{k}f_{01}'(x)+r'(x)+2f(x)=0
\end{eqnarray}
For static fields we have $\dot{r}(x)=0=\dot{n}(x)$. 
Using the Gauss law equation (55) we express the energy density as
\begin{eqnarray}
{\cal E}(x)&=&|d_{x}\chi-r(x)\chi(x)|^{2}+\frac{k}{2}\left(\frac{1}{k}r'(x)
+n(x)-f(x)\right)^{2}
\nonumber\\
&&+\frac{1}{2k}\left\{n'(x)-f_{01}(x)\right\}^{2}+\left\{r(x)f(x)\right\}'
+\frac{1}{k}\left\{n(x)f_{01}(x)\right\}'
\end{eqnarray}
In the case of the static fields we have the following coupled first order 
self-dual equations which can be shown to be consistent with the 
above field equations.
\begin{eqnarray}
&&\frac{1}{k}r'(x)=(f(x)-n(x)),\nonumber\\
&&n'(x)=f_{01}(x),\nonumber\\
&&f'(x)=2r(x)f(x),\\
&&\frac{1}{k}f_{01}'(x)+r'(x)+2f(x)=0\nonumber .
\end{eqnarray}
The energy has a lower of bound zero which is saturated when the fields satisfy 
the above self-dual equations. Eliminating the fields $n,r,f_{01}$ from the 
above equations we get the uncoupled fourth order equation for the field $f$, 
\begin{eqnarray}
\left(-\partial_{x}^{2}+k^{2}\right)\partial_{x}^{2}\ln f
=-4k\left(\frac{1}{2}\partial_{x}^{2}+k\right)f(x)
\end{eqnarray}
which is analogous to the Liouville equation.
The finiteness of energy requires that the field $f$ vanish at both $\pm\infty$.
 It can be easily shown that the fall off is not power law but exponential and 
as $x\longrightarrow -\infty$ we have,
\begin{eqnarray}
f(x)\longrightarrow A e^{Bx}\left(1+Ce^{Bx}\right)+.....
\end{eqnarray}
where A,B and C satisfy the relation
\begin{eqnarray}
-\frac{B^{4}}{k^2}\frac{C}{A}+B^{2}\left(\frac{C}{A}+\frac{2}{k}\right)+4=0
\end{eqnarray}
Using this in the self-dual eqs.(57) and (55) we find
\begin{eqnarray}
n(x)\longrightarrow A_{0}e^{Bx}+.....\nonumber\\
r(x)\longrightarrow \frac{B}{2}+A_{1}e^{Bx}+.....\nonumber\\
f_{01}(x)\longrightarrow A_{2}e^{Bx}+.....
\end{eqnarray}
where the coefficients $A_{0}, A_{1}, A_{2}, B$ and $A$ are related as
\begin{eqnarray}
&&A_{0}=-\frac{A}{\left(1+\frac{2k+B^{2}}{k(k+2)}\right)}\nonumber\\
&&A_{1}=\frac{kA}{B\left(1+\frac{k(k+2)}{2k+B^{2}}\right)}\nonumber\\
&&A_{2}=-\frac{AB}{\left(1+\frac{2k+B^{2}}{k(k+2)}\right)}
\end{eqnarray}
This solution is expected to be symmetric around $x=0$. To know the behaviour
around $x=0$ we can expand the fields around it as 
\begin{eqnarray}
f(x)=\tilde A+\tilde Bx^{2}+\tilde Cx^{4}+...
\end{eqnarray}
Then from eq.(58) we find 
\begin{eqnarray}
k^{2}\left(\tilde{B}+2\tilde{A}^{2}\right)+2k\tilde{A}\tilde{B}-12\tilde{C}=0
\end{eqnarray}
The charge in this case is found to be
\begin{eqnarray}
Q=-\int_{-\infty}^{\infty}dx f(x)=\frac{r(\infty)-r(-\infty)}{2}=\frac{B}{2}
\end{eqnarray}
in case one assumes the same behaviour of the fields at $\pm\infty$.

\section{Conclusion}
In this paper we have studied the domain wall soliton solutions in the 
self-dual Maxwell Chern-Simons systems in $1+1$ dimensions. Here we 
found BPS type bound which is saturated when the self-dual equations holds. 
The numerical solutions for both the topological and the nontopological
soliton equations were obtained. Further the asymptotic properties of the
fields are also studied. We considered the underlying $N=2$ supersymmetry 
of this model by dimensional reduction of the $2+1$ dimensional model. 
Finally we have studied the non relativistic limit of the model. 
This work raises few questions which need to be looked into. For example,
can one generalize the model to non abelian case? 
There may be a gauged version of the nonlinear sigma models with Maxwell
and Chern-Simons term with higher gauge group which could lead to a richer 
variety of two dimensional models with BPS-type energy bound. Further in three 
dimensions, the maximal supersymmetry for Maxwell Chern-Simons system is 
$N=3$ \cite{hckao}. In two dimensions it may be worth enquiring as to what
maximal supersymmetry is allowed. Also 
it would be interesting to study the fermionic and bosonic zero modes
and to check whether the energy bound is saturated at the quantum level
or if there are quantum corrections. 

\section{ACKNOWLEDGMENTS}
I am grateful to Avinash Khare for suggesting this problem and for discussions
as well as for careful reading of the manuscript. Also I thank Pijush K. Ghosh 
and Sanatan Digal for discussions.

%\end{multicols}
\end{document}